\documentclass[prd,tightenlines,nofootinbib,showpacs,preprintnumbers,superscriptaddress, notitlepage, twocolumn ]{revtex4-1}
\usepackage{amsfonts,amsmath,amssymb,amsthm,bbm,hyperref}
\usepackage{graphicx}
\usepackage{color}
\usepackage{comment}
\usepackage[T1]{fontenc}
\usepackage[utf8]{inputenc}

\newcommand{\be}{\begin{equation}}
\newcommand{\ee}{\end{equation}}
\newcommand{\beq}{\begin{eqnarray}}
\newcommand{\eeq}{\end{eqnarray}}
\newcommand{\nn}{\nonumber}

\begin{document}

\title{An approach to study the adiabaticity and irreversibility in the TDHO}
\author{Salvador J. Robles-Pérez}
\email{sarobles@math.uc3m.es}
\affiliation{Departamento de Matemáticas, Universidad Carlos III de Madrid. ROR: https://ror.org/03ths8210,\\ Avda. de la Universidad 30,  28911 Leganés (Madrid), Spain.}
\author{Salvador Castillo-Rivera}
\email{scastill@math.uc3m.es}
\affiliation{Departamento de Matemáticas, Universidad Carlos III de Madrid. ROR: https://ror.org/03ths8210,\\ Avda. de la Universidad 30,  28911 Leganés (Madrid), Spain.}

\date{\today}

\begin{abstract}
This work studies the relationship between parametric amplification (or particle creation), adiabaticity and irreversibility in the non-quasi-static regime of a time-dependent quantum harmonic oscillator (TDHO) that evolves unitarily. We provide analytical results for the evolution of the TDHO valid for any functional value of the frequency, which enables us to monitor the behavior of the thermodynamical magnitudes in the non-quasi-static regime. In the latter, the largest modes of the energy eigenstates commonly undergo a process of spontaneous thermalization, where the concept of temperature naturally arises from the unitary evolution of the oscillator, i.e. without relation to any external source of temperature or thermal bath. As the evolution is unitary, this thermalization process can be reversible. We {extend} the standard definitions of quantum heat and work to account for the change in the populations of the energy levels {and relate it to the diagonal entropy to obtain an effective temperature from the associated second principle of thermodynamics. The qualitative behavior of this temperature is similar to the temperature of the largest modes and is only significant in the non quasi-static region, returning to the original value $T\rightarrow 0$ afterwards. This effect is briefly analyzed in the case of a periodically driven quantum system, where the energy eigenstate mixing and the temperature grow and return to the original value periodically.}

\end{abstract}

\maketitle



\section{Introduction}\label{sec00}

The search for a quantum theory of thermodynamics is one of the most challenging and promising lines of research in contemporary physics. For overall definitions, historical reviews, and prospective analysis, see \cite{Gemmer2009, Tsallis2009, Spakovsky2014, Mahler2015, Vinjanampathy2016, Goold2016, Alicki2018, Deffner2019, Manirul2020}. On the one hand, it is assumed to supersede the classical theory of thermodynamics, and one would expect new non-classical (or semiclassical) thermodynamical effects. On the other hand, it is also expected that a quantum theory of thermodynamics should be derived just from the basic principles of quantum mechanics: quantum (deterministic) evolution and quantum (stochastic) measurement. As pointed out in Ref. \cite{Manirul2020}, that is \emph{one of the big challenges in physics}.

In that search, there are some concepts that are customarily brought together, like the creation of particles ($\dot n \neq 0$) and the non-adiabaticity ($\dot Q \neq 0$) of the system. It occurs in the so-called quasi-static approximation of quantum mechanics, which is paradigmatically represented by a harmonic oscillator with slow varying time dependent frequency, $\omega(t)$. In that case, if $\dot\omega/\omega \ll 1$, the creation of particles induced by  the time-dependent harmonic oscillator is of order $(\dot\omega/\omega)^2$ and, up to  first order the Hamiltonian can be written as
\be\label{Hho01}
H = \hbar \omega(t) \left( \langle\hat n \rangle +\frac{1}{2} \right) ,
\ee
where the number of particles, $\langle\hat n \rangle$, is constant. Due to the evolution driven by the Hamiltonian \eqref{Hho01} is unitary, the conventional definition of quantum heat $Q$ is zero, and the process is called adiabatic. Therefore, the quasi-static approximation is sometimes also called the \emph{adiabatic approximation} \cite{Zeldovich1971, Eliezer1976, Mahler2015, Deffner2019, Zhang2020}.

However, this is not the case in a more general approach to quantum thermodynamics. As it is well-known \cite{Louisell1961, Zeldovich1971, Parker1972, Brown1979, Hu1986, Hu1987, Kandrup1988, Kandrup1989}, the unitary evolution of the time-dependent harmonic oscillator (TDHO) may entail the creation of particles, and in that case, $Q = 0$ but $\dot n \neq 0$.  Furthermore, the quasi-static approximation cannot always be applied, for example, in the so-called finite-time thermodynamics \cite{Geva1992, Hoffman2001, Deffner2019} or in the cycles involved in some quantum thermodynamical machines \cite{Pezzutto2019, Picatoste2024}. It seems interesting to study the quantum thermodynamics of the TDHO at any regime of $\dot \omega$. It might optimize some thermodynamic processes and even open the possibility of new applications.

Two other concepts are usually connected: thermalization and irreversibility \cite{Linden2008, Rigol2008, Spakovsky2014, Vinjanampathy2016, Landi2021}.  {The typical scenario  \cite{Caldeira1983, Joos2003, Schlosshauer2007,  Linden2008, Rigol2008, Landi2021, Picatoste2024}, consists of a central system that  interacts (weakly) with an environment that is \emph{already} in thermal equilibrium at some  temperature $T$. Following the usual approach (see, for instance, Refs. \cite{Joos2003, Schlosshauer2007}), the interaction produces dissipative terms that are responsible for both the increase in the entropy (irreversibility) and the thermalization of the central system to the same temperature as the surrounding environment.

While the above process is very effective in explaining the thermalization process in an open system in contact with a thermal bath, it is not very appropriate for explaining the nature and emergence of the concept of temperature. After all, as noted in Ref. \cite{Deutsch1991}, starting from an environment in thermal equilibrium at temperature $T$ basically shifts the question about the thermalization process of that environment. In order to address that question, we need to study the thermalization that may occur in the evolution of a closed system evolving unitarily \cite{Greiner2002, Gangardt2008, Rigol2008, Cazalilla2011, Bunin2011, Cazalilla2012a, Mori2018}. The main problem in that case is that, strictly speaking, a system that evolves unitarily does not thermalize \cite{Dalessio2016, Mori2018}. If the system is initially in a pure state, under unitary evolution it will remain in a pure state at any later time and not in the diagonal mixed state that is associated with a thermal state.

We can consider, however, effective processes of thermalization \cite{Neumann1929, Cugliandolo1997, Biroli2015, Dalessio2016, Foini2017}. Among the different processes, one approach \cite{Calabrese2007, Sotiriadis2009} consists of studying whether the system evolves in such a way that expectation values of macroscopic observables, when averaged over long periods of time, give the same values that would be obtained from averaged values over the canonical Gibbs ensemble (or, more generally, over the Generalized Gibbs ensemble \cite{Foini2017}). This can be called an effective macroscopic thermalization, or MATE \cite{Mori2018}.

There is also a microscopic concept of thermalization (MITE), a purely quantum notion with no classical analogue \cite{Mori2018}, in which a quantum system is thermalized if the probability distribution of its energy levels follows the usual distribution of a thermal state at a given temperature $T$. This concept of microscopic thermalization is more fundamental in the sense that microscopic thermalization implies macroscopic thermalization, but not necessarily the reverse, which is also the basis of the Eigenstate Thermalization Hypothesis (ETH) \cite{Deutsch1991, Srednicki1994}.

The two previous notions of thermalization are associated to corresponding concepts of temperature, macroscopic and microscopic \cite{Mori2018}. There is still another concept of temperature, which is derived from the second law of thermodynamics,
\be
dS = \frac{d Q}{T} ,
\ee 
which again relates thermalization and entropy \cite{Cugliandolo1997, Polkovnikov2008, Polkovnikov2011, Dalessio2016}. However, the magnitudes involved in this definition of temperature must be revisited in the context of a closed system that evolves unitarily \cite{Polkovnikov2008}, for which the von Neumann entropy is constant, so $dS=0$, and the usual definition of heat also results in $dQ=0$. We need new definitions of entropy and heat in this case \cite{Polkovnikov2008, Polkovnikov2011}.}
The quantum Zeno effect (QZE) is a phenomenon in which the time evolution of a quantum state slows due to systematic measurements of the system. Moreover, frequent measurements may have the opposite effect and speed up evolution under somewhat different conditions. This effect is called the quantum anti-Zeno effect (QAZE) and arises when the frequent measurements are not fast enough \cite{Ghasemi2019}.

The rising interest in the QZE is due to its applications to spin polarization in gases and to the control of decoherence in quantum computing, among others. The physical meaning of decoherence has been characterized by quantum and classical irreversibility, emphasizing that decoherence is an irreversible process as a consequence of the interaction between the quantum system and its environment. Lately, a thermodynamic procedure to irreversibility in quantum systems has been derived based on the continuous interaction between the setting electromagnetic waves and the matter, studying the absorption-emission of a photon by an atomic electron, getting a thermophysical model of quantum thermodynamics in accordance with the observed outcomes. This procedure can be expanded to the QZE, taking into account the coupled quantum system-experimental instrument as the isolated system, while the quantum system and the instrument are independent two open subsystems \cite{Lucia2023a}.

In this setting, "measurements" should be considered as interactions with an external system that disturb the unitary evolution of the quantum system. A slowing of the time evolution, as opposed to an entire freezing, is commonly considered proof of the QZE \cite{Itano2019}. The essential role of irreversibility even appears in QZE. Especially interesting are the thermodynamic relations of quantum computing. Certainly, it is assumed that quantum computing is in principle reversible. In fact, QZE has been shown to play a basic role in the control of the decoherence in the analysis of the Josephson junction in superconducting qubits \cite{Lucia2023b}.

{In this work, we use the relatively simple model of a quantum harmonic oscillator with time dependent frequency that evolves unitarily to address some of these questions. The TDHO offers the main advantage that its quantum state can be described exactly in terms of the invariant representation, for any functional value of the frequency at any time. This allows for compact, analytical solutions that could describe a wide variety of models, from well-isolated systems \cite{Dalessio2016, Cazalilla2012a, Biroli2015}, ultracold atoms out of equilibrium \cite{Langen2015, Santos2011, Mori2018}, time dependent protocols for quantum quenches \cite{Calabrese2007, Kaufman2016, Santos2011, Sotiriadis2009} or periodically driven systems \cite{Grifoni1998, Greiner2002, Kinoshita2006, Bunin2011,  Larson2013, Lazarides2014a, Lazarides2014b, Dalessio2016}. The approach will be eminently theoretical, although some aspects about the experimental implementation will be offered at the end of the article.

The paper is outlined as follows. In Sec. II we describe the exact quantum state of the TDHO in the basis of energy levels of the Hamiltonian. We show that in the non quasi-static regime the largest energy modes follow a thermal distribution with a value of the temperature that naturally emerges from the unitary evolution of the oscillator. In Section III, we review the definitions of heat and entropy for the case of a closed system that evolves unitarily and, for the TDHO, we calculate the macroscopic temperature derived from the second law of thermodynamics. Finally, we analyze the possible experimental implementation of the model and draw some conclusions in Section IV.
}


\section{Evolution of the TDHO: reversible thermalization}\label{sec01}

{ Let us consider an unspecified experiment modeled by a time dependent harmonic oscillator (TDHO). As a guiding reference, one can consider an experimental setup with ultracold atoms, which can be considered a good example of highly controllable isolated quantum systems \cite{Langen2015, Mori2018}. However, we will not restrict our attention to a particular experiment because one of the advantages of  the TDHO is that it can model many different ones. The time dependence of the frequency indicates that the system is not truly isolated. However, as considered in Refs. \cite{Bunin2011, Biroli2015}, one can assume that the coupling to external dissipative degrees of freedom is strongly suppressed  on accessible timescales and  that the effects of the environment are condensed in additional self energy terms, which in the case of the harmonic oscillator are determined by the different values of the frequency. Moreover, the time dependence allows us to monitor dynamical processes, which is needed to derive thermodynamical relations \cite{Dalessio2016}. In particular, we will assume that these dynamic processes originate from the change of macroscopic parameters of the Hamiltonian \cite{Dalessio2016} without breaking the unitary evolution of the model.}

The quantum state $ |\psi \rangle$ of a harmonic oscillator with time-dependent frequency\footnote{The extension to a harmonic oscillator with time-dependent mass as well is straightforward.}, $\omega(t)$, is given by the solution of the Schrödinger equation
\be\label{sch01}
i\hbar \frac{\partial}{\partial t} |\psi(t) \rangle = \hat H_S | \psi(t) \rangle ,
\ee
with
\be\label{ham01a}
\hat H_{S} = \frac{1}{2} \hat p_{x, S}^2 + \frac{\omega^2(t)}{2} \hat x_S^2 ,
\ee
where $\hat x_S$ and $\hat p_{x, S}$ are the position and momentum operators, and the subscript $S$ in the Hamiltonian \eqref{ham01a} means that we are working in the Schrödinger picture, where the operators $\hat p_{x,S}$ and $\hat x_S$ do not depend on time and the time dependence is enclosed in the state, $|\psi(t)\rangle$. As it is well known \cite{Lewis1968, Lewis1969, Leach1977, Leach1983, Hartley1982, Ray1982, Pedrosa1987, Dantas1992, Kanasugui1995, Seleznyova1995, Yeon1997, Song2000, Park2004}, the solution of the Schrödinger equation \eqref{sch01} can be written in terms of a complete set of wave functions, $\{\psi_N(x,t)\}_{N \in \mathbb N}$, which are the eigenfunctions of an invariant operator associated to the Hamiltonian \eqref{ham01a}. The basic idea is that there is a unitary transformation that relates the Hamiltonian of the TDHO \eqref{ham01a} with the Hamiltonian of the harmonic oscillator with constant frequency, $\omega_0$. The same transformation relates the solutions of their corresponding Schrödinger equations. In that case, we can obtain a complete orthonormal set of solutions of the Schrödinger equation of the TDHO from the complete orthonormal set of solutions of the Schrödinger equation of the time-independent harmonic oscillator. The orthonormal set of solutions of the Schrödinger equation \eqref{sch01} turns out to be \cite{Leach1983, Kanasugui1995}
\be\label{WFInv01}
\psi_N(x,t) =  \frac{e^{-i \left( N +\frac{1}{2}\right) \omega_0 \tau(t)}}{\sqrt{2^N N! \sigma} }\left( \frac{\omega_0}{\pi} \right)^\frac{1}{4} e^{\left(i\frac{\dot \sigma}{2\sigma} -\frac{\omega_0}{2\sigma^2}\right)  x^2} \text{H}_N(\frac{\sqrt{\omega_0}x}{\sigma})
\ee
where $\omega_0=\omega(0)$, the function $\tau(t)$ is given by
\be
\tau(t) = \int^t \frac{1}{\sigma^2(t')} dt'
\ee
and $\sigma(t)$ is the solution of the auxiliary equation
\be\label{sigma01}
\ddot\sigma +\omega^2(t) \sigma = \frac{\omega_0^2}{\sigma^3} ,
\ee
satisfying the initial conditions, $\sigma(0)=1$ and $\dot\sigma(0) = 0$. With these initial conditions, the wave function  \eqref{WFInv01} is at time $t=0$ the wave function of the number states of the harmonic oscillator with constant frequency $\omega_0$.

{The Ermakov equation \eqref{sigma01}, and Lewis-Riesenfeld invariants allow us to figure out exact solutions to the time-dependent Schrödinger equation for systems with time-dependent hamiltonians \cite{kim2016, chen2011}}. The quantum state $|\psi(t)\rangle$ in Eq. \eqref{sch01} can be written as, 
\be
|\psi(t)\rangle = \sum_M C_M |M(t)\rangle ,
\ee
where $C_M$ are constant coefficients. In particular, if the TDHO is initially in the state $|N\rangle \equiv |N(0)\rangle$, then $C_M= \delta_{MN}$ and the TDHO will remain in the state $|N(t)\rangle$ along the entire evolution of the harmonic oscillator. However, the number states $|N(t)\rangle$ represented by the wave functions \eqref{WFInv01} are not the energy eigenstates of the Hamiltonian \eqref{ham01a}, $|N_\omega(t)\rangle$, which are instead represented  by the wave functions
\be\label{WFOme01}
\psi_N^{(\omega)}(x,t) =  \frac{e^{-i \left( N +\frac{1}{2}\right) \omega(t)}}{\sqrt{2^N N! } }\left( \frac{\omega(t)}{\pi} \right)^\frac{1}{4} e^{ -\frac{\omega(t)}{2}  x^2} \text{H}_N(\sqrt{\omega(t)}x)
\ee
{ The choice of the basis that is used to represent the quantum state of the system becomes a subtle issue \cite{Dalessio2016} because the parametric creation of particles and the related thermodynamical magnitudes are caused by the mixing of the different basis states along the evolution of the system \cite{Jensen1985, Calabrese2007, Larson2013, Lazarides2014a, Foini2017}, and the mixing clearly depends on the choice. Throughout this work, following Refs. \cite{Polkovnikov2008, Rigol2008, Polkovnikov2011, Dalessio2016}, we will assume that the basis of instantaneous energy eigenstates is the basis that physically represents  the outcomes of the measurement in the experiment. However, it is worth point out that other representations could be used\footnote{{For instance, in cosmology \cite{Hu1986, Hu1987, Kandrup1988, Kandrup1989} it is customary to choose the basis of eigenstates of the Hamiltonian at some initial time (the remote past, for instance). Eventually, the choice is rooted in the projection postulate of quantum mechanics, the measurement process and the concept of particle, which are all complex and fundamental issues in the quantum theory.}} and that the thermodynamical description may eventually depend on the choice.}

In terms of these energy eigenstates, the evolution of an initial number state $|N\rangle$ can be written at time $t$ as
\be
|N(t)\rangle = \sum_M C_{MN}^{(\omega)}(t) \, |M_\omega(t) \rangle ,
\ee
where the time dependent coefficients, $C_{MN}^{(\omega)}(t) = \langle M_\omega(t) | N(t)\rangle$, can be calculated in terms of the associated Legendre functions $P_\mu^\nu(z)$ (see App. \ref{MElements01} and \cite{Yuen1976, Brown1979, Kim1989a}),
\be\label{RhoMN0_01}
C_{MN}^{(\omega)}(t) =  (-1)^\frac{M-N}{2} \sqrt{\frac{M!}{N!}}  \frac{e^{i\varphi}}{\sqrt{|\alpha_\omega|}}P^\frac{N-M}{2}_\frac{N+M}{2}\left(\frac{1}{|\alpha_\omega|}\right)
\ee
if $M\pm N$ is an even integer\footnote{This is a consequence of the fact that the particles are created (or destroyed) in pairs in the parametric amplifier \cite{Parker1977, Hu1987, Kandrup1988, Kim1989}.} and zero otherwise, with $i\varphi$  a pure imaginary phase given by,
\be
i\varphi = \frac{i\theta_\alpha}{2}(N+M+ 1)+\frac{i\theta_\beta}{2}(M-N) ,
\ee
{where $\theta_\alpha = \arg(\alpha_\omega)$ and $\theta_\beta = \arg(\beta_\omega)$, with $\alpha_\omega$ and $\beta_\omega$ the coefficients of the Bogolyubov transformation that relates the diagonal representation at time $t$ with the initial representation,
\beq\label{alpha01}
\alpha_\omega(t) &=& \frac{1}{2} \sqrt{\frac{\omega(t)}{\omega_0}} \left( \sigma + \frac{\omega_0}{\sigma \omega(t)}  + \frac{i\dot \sigma}{\omega(t)} \right) \, e^{-i\omega_0\tau} \\ \label{beta01}
\beta_\omega(t) &=& \frac{1}{2} \sqrt{\frac{\omega(t)}{\omega_0}} \left( \sigma - \frac{\omega_0}{\sigma \omega(t)}  + \frac{i\dot \sigma}{\omega(t)} \right) \, e^{+i\omega_0\tau} ,
\eeq
with, $|\alpha_\omega|^2-|\beta_\omega|^2 = 1$. There is in general a mixing\footnote{{Mixings can be found whenever we use two different representations related by a Bogolyubov transformation (see, for intance, Refs. \cite{Calabrese2007, Foini2017}).}}  of the energy eigenstates along the evolution of the TDHO or} a transition from the initial number state, $|N\rangle$, to the energy eigenstates $|M_\omega(t)\rangle$ that is time dependent along the evolution of the TDHO. The probability of measuring the initial number state $|N\rangle$  in the energy eigenstate $|M_\omega(t) \rangle$ at time $t$ is given by 
\be\label{ProbOme_01}
P^{(\omega)}_M(N;t) =\frac{M!}{N!} \frac{1}{|\alpha_\omega|} \left| P^\frac{N-M}{2}_\frac{N+M}{2}\left(\frac{1}{|\alpha_\omega|}\right) \right|^2 .
\ee
In the quasi-static approximation, $|\alpha_\omega| = 1 + \mathcal O(\frac{\dot\omega}{\omega})$ and, $P^{(\omega)}_M(N;t)\approx \delta_{MN}$. Nevertheless, in the non-quasi-static regime of the frequency, there is mainly a time-dependent evolution of the probabilities of measuring the TDHO in the energy eigenstates, given by Eq. \eqref{ProbOme_01} { (see, Fig. \ref{fig_Pomega02})}.

\begin{figure}

\includegraphics[width=7cm]{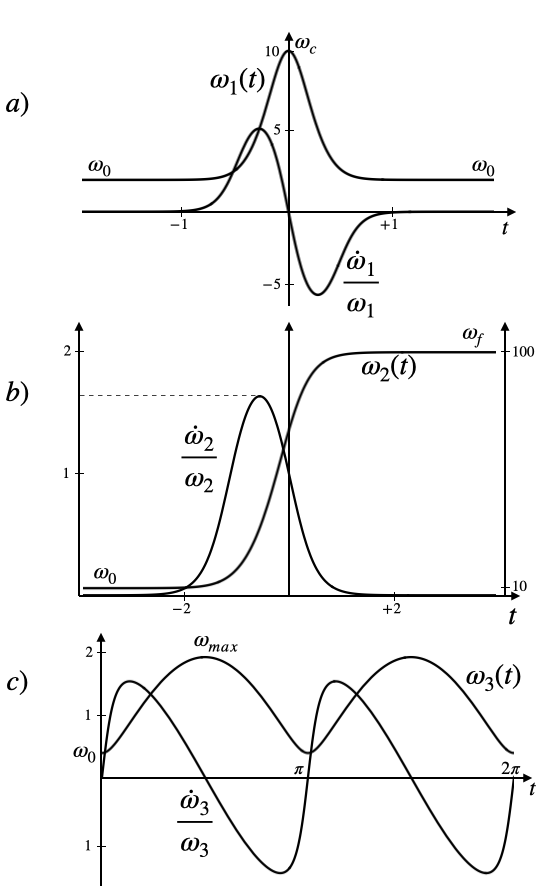}

\caption{Frequency $\omega(t)$ and its time derivative for three different cases: $a)$ the frequency $\omega_1(t)$ starts at $\omega_0$ in the remote past and approaches $\omega_f$ in the remote future, with a central region of transition; $b)$ $\omega_2(t)$ starts at the value $\omega_0$, it reaches the value $\omega_c$ at the central non quasi-static region, and returns to the original value; and, $c)$ the (Floquet) frequency, $\omega_3(t)$, periodically oscillates  between the values $\omega_0$ and $\omega_{max}$. The regions for which $\left | \frac{\dot\omega}{\omega}\right | \not\ll  1$ are the non  quasi-static regions.}

\label{fig_freq01}

\end{figure}

\begin{figure}

\includegraphics[width=7cm]{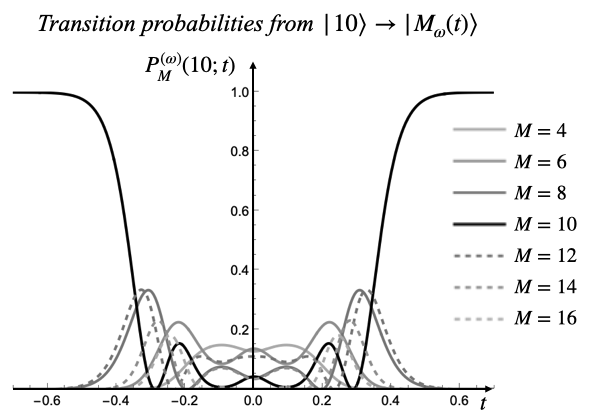}

\caption{Probability, $P_M^{(\omega)}(1;t)$, given by \eqref{ProbOme_01}, of finding the TDHO in the energy eigenstate $|M_\omega(t)\rangle$, for $\omega=\omega_2(t)$ of Fig.  \ref{fig_freq01}, when the initial state is the number state $|10\rangle$. The numerical values are those used in Fig. \ref{fig_freq01}. The mixing of energy eigenstates only occurs in the non quasi-static central region.}

\label{fig_Pomega02}

\end{figure}

\begin{figure}

\includegraphics[width=6.5cm]{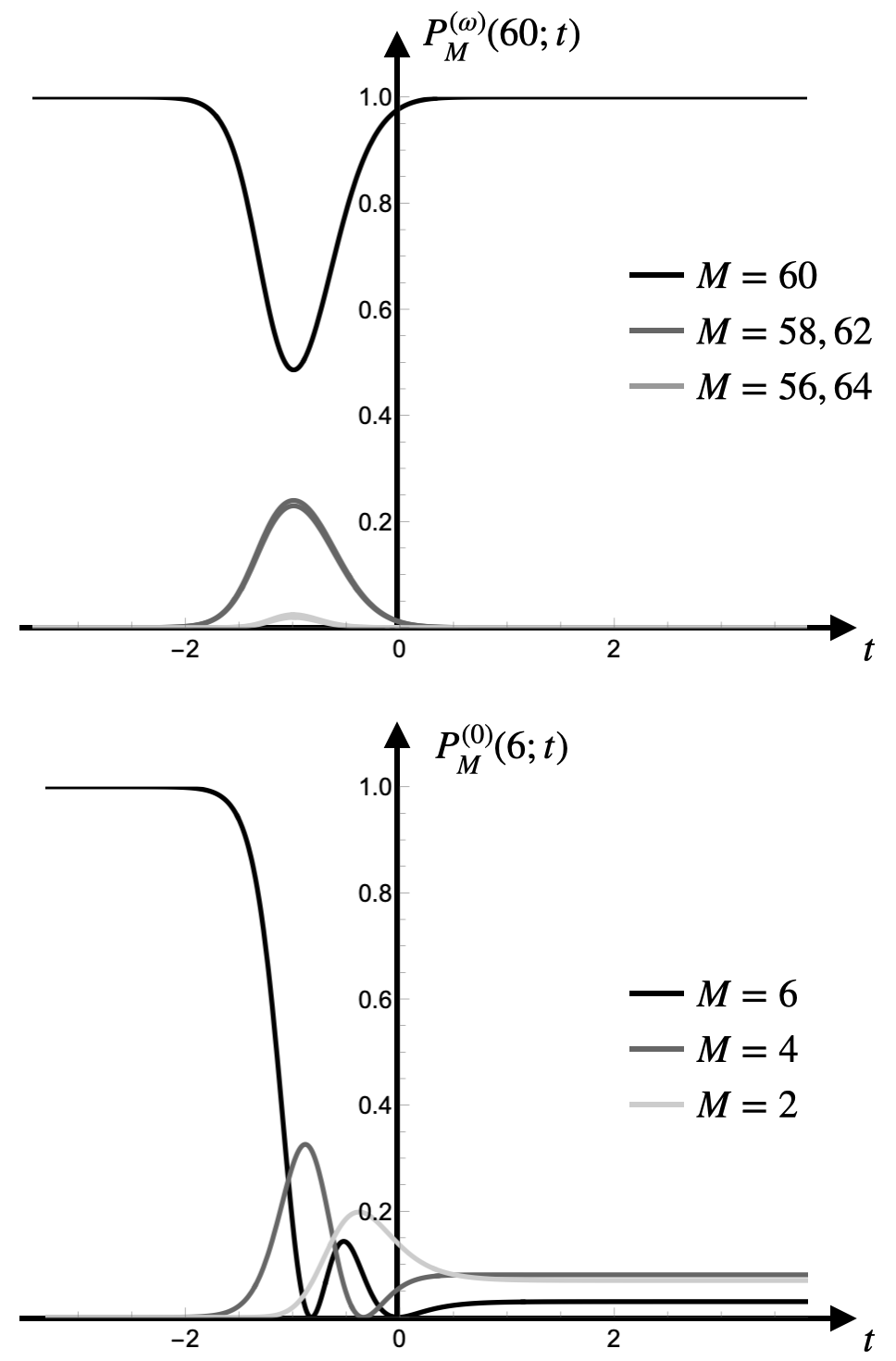}

\caption{Difference between transition probabilities when calculated in terms of the energy eigenvalues (up) or in terms of the initial eigenstates (down), for $\omega=\omega_1(t)$ in Fig. \ref{fig_freq01}. The mixing of energy eigenstates only occurs in the non quasi-static central region, after which the TDHO returns to the initial number eigenstate but of the final Hamiltonian, with $\omega_f$. In terms of the initial representation, the mixing depends on the difference between the initial frequency, $\omega_0$, and the value of $\omega(t)$, so there is mixing of initial eigenstates even at the final quasi-static region of $\omega_1(t)$ (see, Fig. \ref{fig_freq01}).}

\label{fig_comparative}

\end{figure}

Let us now analyze the quantum state of the TDHO from the point of view of the density matrix operator, $\hat \rho_S$, which can describe pure and mixed states \cite{Joos2003, Schlosshauer2007, Jaeger2007, Gemmer2009}. If the harmonic oscillator is initially prepared in a number state $|N\rangle$, the density matrix at time $t$ can be written in the energy eigenstate basis as 
\be\label{DM01}
\hat\rho_S(t) =  \sum_{I, J} P^{(\omega)}_{I,J}(N;t) | I_\omega(t) \rangle \langle J_\omega(t) | ,
\ee
where,
\beq\label{Pijome02}
P^{(\omega)}_{I,J}(N;t) = \langle I_\omega(t) | N(t)\rangle \langle N(t) | J_\omega(t)\rangle .
\eeq
We can split the density matrix equation \eqref{DM01} into the diagonal and the non diagonal terms,
\be\label{rhodiag02}
\hat \rho_S =  \sum_{I} P^{(\omega)}_{I}(N;t) | I_\omega\rangle \langle I_\omega | +  \sum_{I\neq J} P^{(\omega)}_{I,J}(N;t) | I_\omega\rangle \langle J_\omega | ,
\ee
where, for simplicity, we have removed the explicit time dependence on the energy eigenstates. The probabilities $P^{(\omega)}_I(N;t)$ are given by Eq. \eqref{ProbOme_01}, and the off-diagonal elements are given by Eq. \eqref{Pijome02}. The diagonal terms of the density matrix are the probabilities of measuring the TDHO in the corresponding energy eigenstate at time $t$. The off-diagonal terms measure the non-classical correlations between them. We can show that the largest modes of the energy eigenstates approach a thermal distribution along the evolution of the TDHO. Let us note that for the largest modes, using Eq. (8.755.1) of Ref. \cite{Gradshteyn2007}, the associated Legendre functions can be approximated by
\be\label{ALapp01}
P^\frac{N-I}{2}_\frac{N+I}{2}\left(\frac{1}{|\alpha_\omega|}\right) \approx  P^\frac{-I}{2}_\frac{+I}{2}\left(\frac{1}{|\alpha_\omega|}\right) = \frac{1}{\left(\frac{I}{2}\right)!} \left( \frac{|\beta_\omega|}{2 |\alpha_\omega|} \right)^\frac{I}{2} ,
\ee
where { $\alpha_\omega$ and $\beta_\omega$ are given by \eqref{alpha01} and \eqref{beta01}, respectively}. By using \eqref{ALapp01} and Stirling's approximation, the diagonal part of the density matrix \eqref{rhodiag02} can be rewritten as\footnote{In the sum \eqref{rhodiag03}, the quantum number $I$ has a definite parity, i.e. it is either an even or an odd number.}
\beq\nn
\hat \rho &=&  \frac{1}{N!} \sqrt{\frac{2}{\pi}} \frac{1}{\cosh r_\omega} \sum_{I} \left( \frac{\tanh r_\omega}{I^\frac{1}{2I}} \right)^{I} | I_\omega \rangle \langle I_\omega | +  \ldots , \\ \label{rhodiag03}
&=&   \frac{2}{Z } \sum_{I} C_I \, e^{-\frac{\omega(t)}{T_I(t)} \left( I+\frac{1}{2} \right)} | I_\omega\rangle \langle I_\omega | +  \ldots , 
\eeq
where the ellipses in Eq. \eqref{rhodiag03} include the lowest modes and the off-diagonal terms, which are not considered 
\beq
Z &=& N! \sqrt{\pi \sinh 2r_\omega } , \\
C_I &=& I^\frac{1}{4 I} , \\ \label{TK0_01}
T_I(t) &=& \frac{\omega(t)}{ \log (\coth r_\omega) + \frac{1}{2 I} \log I} ,
\eeq
where, $\sinh r_\omega = |\beta_\omega|$ and $\cosh r_\omega = |\alpha_\omega|$. $T_I(t)$ cannot be considered a proper temperature because it depends on the value of the mode $I$. However, the resemblance of the state in Eq.\eqref{rhodiag03} with a thermal distribution is relevant. In fact, in the limit $I\rightarrow \infty$, $C_I \rightarrow 1$, and
\be\label{T_001}
T_I(t) \rightarrow T(t) = \frac{\omega(t)}{ \log (\coth r_\omega)} = \frac{\omega(t)}{ \log \frac{|\alpha_\omega|}{|\beta_\omega|}}  .
\ee
In the non-quasi-static regime, there is then a process of thermalization of the highest (most classical) modes of the energy eigenstates along the unitary evolution of the TDHO. The temperature in Eq. \eqref{T_001} does not depend on the specific conditions of the environment and, particularly, it is not related to any external bath, but it naturally emerges as a property of the non-quasi-static evolution of the TDHO in the energy eigenstates basis.
{ 
The strong-field laser phenomenon, high-harmonic generation (HHG), was originally observed in gases between the 1980s and 1990s. The exploration of HHG has extended to solids. HHG in solids has enabled applications in several parts of electronic band structure, topological states of matter, phase transitions, etc. HHG is commonly explained in terms of atomic gases.  The semiclassical method, in which light is treated classically and matter quantum mechanically, takes into consideration oscillations of the dipole of a field-induced bound-continuum state.  It has a three-step explanation: an electron is ionised by a strong laser field, after that propagates within the field-dressed continuum and recombines with an ionic core.  It transforms its excess energy into ultrafast pulses of light with odd frequency multiples of the driving laser’s frequency.  An analogous process has been researched in solids, known as the interband mechanism. There exists an intraband mechanism in solids, in addition to the interband. The first one characterises the generation of harmonics as a response to charge movement within individual electronic bands \cite{Thorpe2025}.
The HHG spectrum is composed of a set of frequencies showing the predicted exponential decrease in efficiency for the low-order harmonics (perturbative regime), observed by an almost constant plateau for the high-order odd harmonics (nonperturbative regime) that terminates suddenly at some cut-off frequency \cite{Moreno2023}. Liu et al. \cite{Liu2017} have studied the wavelength scaling of the cutoff energy for the solid HHG. They split the contributions into intraband and interband elements. As a result, the cutoff energy of the intraband HHG was wavelength independent. On the other hand, the cutoff energy of the interband HHG displayed a linear dependence on laser wavelength. Particularly, the cutoff energy of the detected harmonics at all times is linearly conditional on the field strength, since both the intraband and interband HHG establish a linear cutoff rule on the field strength. Thorpe et al. \cite{Thorpe2025} presented a quantum optical extension of the quantum-matter Lewenstein model of HHG in gases, which involves two channels: the interband and intraband HHG in solids.  In gases, like in solids, intraband HHG commands at low orders. In particular, they have researched HHG in argon gas driven by a strong linearly polarised laser field with a Gaussian temporal envelope. The total photon yield is computed as a function of harmonic order using driving wavelengths of $1.05$ $\mu m$, $2.1$ $\mu m$, $3.05$ $\mu m$, and $4.2$ $\mu m$, mainly. In the model, the lowest order harmonics are driven by the intraband, with intraband dominance rising with wavelength. The intraband dominance noticeably occurs for more harmonics as the wavelength is raised. As can be seen, the wavelength has a relevant role in the harmonic order, and the harmonic limit changes according to the corresponding values of wavelength.  

}
{ At this point, the following observations are appropriate:

\emph{1}. In the representation of the energy eigenstates, there is only mixing of states in the non quasi-static regime, where $\frac{\dot\omega}{\omega} \gtrsim 1$. When the frequency re-enters a quasi-static regime zone, in which $\frac{\dot\omega}{\omega} \ll 1$, the probabilities are re-established to the original values, i.e. $|\alpha_\omega| \rightarrow 1$ (see comments after Eq. \eqref{ProbOme_01}) and the quantum state of the TDHO returns to a pure state, with $T\rightarrow 0$. It does not matter whether the final value of the frequency equals the original value (see the different cases of frequencies $\omega_1(t)$ and $\omega_2(t)$ in Fig. \ref{fig_freq01}). If at time $t=0$ the TDHO starts in an energy eigenstate of the Hamiltonian, $|N_{\omega_0}\rangle$, then after passing through the non-quasi-static zone, the TDHO returns to the pure state $|N_{\omega_f}\rangle$, regardless of whether $\omega_f = \omega_0$ or not (see, Fig. \ref{fig_Pomega02} and Fig. \ref{fig_comparative} (up)). This is an effect of the energy state representation. If, for instance, we had used the initial representation to describe the evolution of TDHO, the corresponding Bogolyubov coefficients, $\alpha_0$ and $\beta_0$, would not depend on the value of $\frac{\dot\omega}{\omega}$ but on the difference $|\omega(t)-\omega_0|$ at time $t$, and in that case, state mixing could occur even in the final quasi-static zone, provided that $\omega_f \neq\omega_0$ (see Fig. \ref{fig_comparative} (down))

\emph{2}. The fact that only the higher modes of the energy eigenstates of the TDHO thermalize indicates that the quantum state \eqref{rhodiag03} is not truly a thermal state. It can be seen as a realization of the weak ETH \cite{Deutsch2018, Mori2018}, where only a part of the eigenmodes is required to be in thermal distribution. It is true that, strictly speaking, thermalization requires the strong ETH \cite{Deutsch2018}. But as with thermalization in entangled states \cite{Cardy2008}, the fact that part of the quantum state thermalizes even when the quantum state of the whole system remains in a pure state shows that these thermalization processes in isolated, unitary evolving systems depend on how we measure the system. It also depends on the choice of basis, which ultimately represents the measurement process through the projection postulate of quantum mechanics. More specifically,  the statistical behavior of an isolated quantum system depends on the three interrelated factors \cite{Jensen1985}: the initial state, the observable (or the measurement process), and the Hamiltonian  (see, also, Refs. \cite{Calabrese2007, Larson2013, Lazarides2014a, Mori2018}).

\emph{3}. It is also worth noticing that the mode dependence of $T_I$ in \eqref{TK0_01} is not the mode dependence usually associated to the momentum distribution in a field theory \cite{Calabrese2007, Sotiriadis2009, Gangardt2008, Cazalilla2012a, Foini2017, Garriga2005}. In these works, the average value of a certain observable is calculated, and the result is compared with the result that would be obtained by calculating the average value of that observable if the system were in a thermal distribution with a temperature $T$. From this comparison, a macroscopic effective temperature is obtained, which turns out to be independent for each mode because in a free field theory the field modes do not interact with each other \cite{Sotiriadis2009}.

The developments in this work can be easily extended to the description of a free field theory with mass $m$, since in quantum field theory the field is eventually described by a set of harmonic oscillators with a mode dependent frequency given by,
\be
\omega_k^2(t) = k^2 + m^2 ,
\ee
where, $k = |\vec k|$, is the momentum associated to the Fourier component of the field, and the mass $m$ can be extended to a time-dependent value if we want to describe, for instance, dynamical quantum quenches or other protocols (for quantum quenches, see for instance Refs. \cite{Calabrese2007, Sotiriadis2009, Santos2011, Kaufman2016}). In a free field theory, each mode evolves independently and the microcanonical state of the field would be given by
\be
\hat \rho = \prod_k \hat \rho_k
\ee
where the diagonal elements of each $\hat \rho_k$ would be given by a expression similar to \eqref{DM01} for each individual mode. However, the $I,J$-dependance disappears in the average value of any macroscopic observable, which for a given $k$-mode reads
\be\label{ensemAve01}
\langle \hat O \rangle_k = \sum_{I,J} P^{(\omega_k)}_{I,J}(N;t) \langle J_{\omega_k}(t) | \hat O| I_{\omega_k}(t) \rangle  .
\ee
One can then compare the result obtained in \eqref{ensemAve01} with the result that is obtained if we average the operator $\hat O$ with a thermal state in the basis of energy eigenstates at time $t$, and from the comparison, obtaining an effective macroscopic temperature, $T_k(t)$, for each $k$-mode.  But this $k$- dependent effective macroscopic temperature obtained from averaged values of observables is not the $I$-mode dependent microscopic temperature-like parameter $T_I$ given in \eqref{TK0_01}, which is not actually an actual temperature unless we can split the higher from the lowest modes of the TDHO along the evolution.

\emph{4}. If we could separate the higher modes and thus approximate the state of those modes by a thermal state with the temperature $T(t)$, given by \eqref{T_001}, it is worth noticing that the time dependence of $T(t)$ would not break the condition of (quasi-) thermal equilibrium provided that relaxation time is fast compared to time where the energy changes significantly \cite{Biroli2015}. Let us note that the concept of equilibrium is a time scale concept that depends on the observation time scales so one system can be at equilibrium at one scale and not at another \cite{Cugliandolo1997}. Following Ref. \cite{Cugliandolo1997}, the temperature $T(t)$ of the highest energy modes, if separated from the lowest modes, would deserve the name of temperature because: (i) it is the one measured on that sub-system by a thermometer\footnote{Let us recall that, from the ETH, microscopic thermalization implies macroscopic (or observable) thermalization.} whose reaction time is smaller than the relevant time scale of energy change, (ii) it determines the direction of heat flow (see, Sec. III), and (iii) it acts as a criterion for thermalization.

Finally, to obtain a truly thermal state, we need the off-diagonal terms of the density matrix \eqref{DM01} to be suppressed, so  it becomes diagonal. Let us notice that the coefficients $P^{(\omega)}_{I,J}(N;t)$ of the off-diagonal terms of the density matrix operator  are proportional to a phase $e^{i\theta (I-J)}$, which is a general result \cite{Reimann2008, Rigol2008, Rigol2012}. For far off-diagonal elements, these terms become oscillating functions that yield vanishing terms when averaged over sufficiently large time intervals. Then, the observation time scale needs to be larger enough to contain several cycles of the off diagonal phases but small enough with respect the energy  change in the Hamiltonian. That clearly introduces a difficulty in the experimental implementation. However, if these correlations are not destroyed, because for instance we do not perform a measurement on the system, and the complete set of modes can be reassembled (or if they have never been separated), then the process is reversible as long as the evolution of the TDHO is unitary (on relevant time scales).

Recently, quantum decoherence has been modeled using quantum Brownian motion for a particle in a Pöschl-Teller potential interacting with a setting of non-interacting harmonic oscillators in thermal equilibrium \cite{Negarandeh2024} and taking into consideration that a expression for Pöschl-Teller potential is given by $V(x)\propto 1/\cosh^2(x)$ \cite{arias2003, cevik2016}.
These effects can better be seen in specific examples. We will focus on the frequency $\omega_1(t)$ of Fig. \ref{fig_freq01}, which is a frequency that starts at the initial frequency $\omega_0$, then grows along a non-quasi-static central region to the value $\omega_c$, at time $t=0$, and then returns to the original value. It is given by\footnote{The numerical values are, $\omega_0 = 2$, $\omega_c = 100$, and $\kappa = 2$}
\be\label{freq01}
\omega_1^2(t) = \omega_0^2 +  \frac{\omega_c^2 -\omega_0^2}{\cosh^2(\kappa t)} ,
\ee
with an associated solution of the auxiliary equation \eqref{sigma01} satisfying the appropriate initial conditions\footnote{In the case of the frequency \eqref{freq01} the initial conditions are imposed on the remote past, $t \ll 0$.}  
\be\label{sigma101}
\sigma_1(t) =\mathcal{N}_0 \, |P_\nu^\mu(\tanh(\kappa t))| ,
\ee
where $\mathcal N_0$ is a normalization constant, $\mu = \frac{i\omega_0}{\kappa}$, and
\be
\nu = \frac{1}{2}\left( \sqrt{1+\frac{4}{\kappa^2} (\omega_c^2 -\omega_0^2)} - 1 \right) .
\ee
The two other frequencies, $\omega_2(t)$ and $\omega_3(t)$ in Fig. \ref{fig_freq01}, will only be used to analyze specific aspects. Frequency $\omega_2(t)$ will allow us to show the dependence of the transition probabilities and the mixing of eigenstates depending on whether the projection is made onto the instantaneous energy eigenstates at time $t$ or onto the initial energy eigenstates (the energy eigenstates at time $t=0$). It is given by\footnote{The numerical values used in calculations are $\omega_0 =10$, $\omega_f=100$, and $\kappa = 2$.}
\be\label{freq01}
\omega_2^2(t) = \frac{\omega_f^2 +\omega_0^2}{2} +  \frac{\omega_f^2 -\omega_0^2}{2}\tanh \kappa t  ,
\ee
with associated $\sigma$ function given in terms of the hypergeometric function ${}_2F_1(\alpha, \beta, \gamma, z)$ as
\be
\sigma_2(t) = |{}_2F_1(\frac{i \omega_+}{\kappa}, 1+ \frac{i \omega_+}{\kappa} , \frac{i \omega_0}{\kappa}, \frac{1}{2}(1+\tanh\kappa t) )| ,
\ee
where $\omega_+=(\omega_f+\omega_0)/2$. Finally, frequency $\omega_3(t)$ is a Floquet frequency \cite{Strang2005},
\be\label{freq03}
\omega_3^2(t) = a - 2 q \cos 2t ,
\ee
with associated $\sigma$ solution given by
\be
\sigma_3(t) = \sqrt{ x_1^2(t) + C_1^2 \, x_2^2(t) } ,
\ee
where $x_1(a,q;t)$ and $x_2(a,q;t)$ are the normalized Mathieu cosine and sine functions\footnote{The numerical values used in calculations are $a=1.85$ and $q=0.85$.}, respectively, and \cite{Strang2005}
\be
C_1^2 = \frac{1-x_1^2(\pi)}{x_2^2(\pi)} 
\ee

}

{As we have said, let us first focus on the case of a TDHO with frequency $\omega = \omega_1(t)$.} We can use the  function \eqref{sigma101} in \eqref{alpha01} and \eqref{ProbOme_01} to represent the evolution of an initial number state, $|N\rangle$, in terms of the different energy eigenstates, $|N_\omega(t)\rangle$. {An example of the} probability \eqref{ProbOme_01} is represented in Fig. \ref{fig_Pomega02} for an initial value of the quantum number $N=10$. One can see that the redistribution of the population of the energy eigenstates is only significant in the non quasi-static region around $t=0$, as expected. After the non quasi-static region the { energy eigenstate mixing} disappears and the TDHO returns to the original initial state. {The mixing of the energy eigenstates depends on the quantity $\dot\omega/\omega$, and thus it does not matter if the value of the frequency in the quasi-static region after the central quench has the same value as the original frequency (e.g. $\omega_1(t)$ in Fig.  \ref{fig_freq01}) or a different one (e.g. $\omega_2(t)$ in Fig.  \ref{fig_freq01}). This is an effect of the instantaneous energy eigenstate basis. If instead we had considered the basis of the initial energy eigenstates to be that on which the TDHO state is projected, then the mixing of the basis states would depend on the difference $\omega(t)-\omega_0$, and therefore there would be mixing even in the final quasi-static region.(see, Fig. \ref{fig_comparative})}

\begin{figure}

\includegraphics[width=7cm]{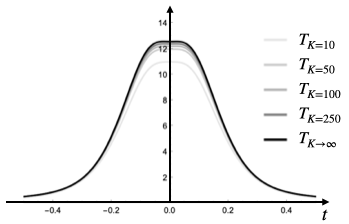}

\caption{Evolution of the mode temperature $T_K$, given by \eqref{T_001}, of the probability distribution of energy eigenstates. The thermal distribution formed by the largest states vanishes after crossing the central region non quasi-static region. }

\label{fig_reversibletemperature}

\end{figure}

{Returning to the representation of instantaneous energy eigenstates, which is the representation considered in this and other  works \cite{Polkovnikov2008, Rigol2008, Polkovnikov2011, Dalessio2016}, the mixing of states and therefore the thermalization of the higher modes of the density matrix disappears after the central non quasi-static zone (see, Figs. \ref{fig_Pomega02} and \ref{fig_reversibletemperature}).}

It is significant to notice that this is an exclusive effect of the evolution of an initial number state in the non-quasi-static regime of the TDHO. In the quasi-static regime, $|N(t)\rangle \approx |N_\omega(t)\rangle$,  and an initial number state $|N\rangle$ evolves as the energy eigenstate $|N_\omega(t)\rangle$, there is no thermalization of the state of the TDHO unless the initial state is already in a thermal state. In that case, the initial state is in a thermal state at temperature $T_0$, the diagonal terms of the density matrix operator in equation \eqref{DM01} can be calculated following the steps given by Ref. \cite{Brown1979}. In the limit $T_0 \rightarrow 0$ (an initial vacuum state), one obtains for the diagonal elements $P^{(\omega)}_{I}(t)$ a thermal distribution with the temperature in equation \eqref{T_001}, as expected. In the opposite limit, if the initial temperature $T_0$ is large enough compared with the parametric amplification caused by the time dependence of the frequency, or more specifically, when the initial value of the temperature satisfies
\be
4 |\alpha_\omega|^2 |\beta_\omega|^2 \sinh^2 \frac{\omega_0}{T_0 } \ll 1 ,
\ee 
a thermal distribution of the diagonal element of the density matrix operator is obtained with a temperature that can be approximated by
\be\label{T05}
T(t) \approx \frac{\omega(t)}{\omega_0} T_0 \left( 1 - 2 |\beta_\omega|^2\right)
\ee
The first term in equation \eqref{T05} is the time-dependent evolution of the temperature of a thermal state that evolves quasi-statically from the initial temperature $T_0$ \cite{Kim2022}. The second term is a first-order correction to the quasi-static evolution of the temperature, which is of order $\left( \frac{\dot\omega}{\omega}\right)^2$. The general effect of a time-dependent frequency in a pre-existing thermal state is the amplification or the reduction (for instance, after crossing the non-quasi-static region around $t=0$) of the initial temperature, $T_0$. It occurs only when the initial state is a pure number state evolving in a non-quasi-static regime, when the temperature becomes an emergent feature of the quantum evolution of the TDHO.


\section{Adiabaticity and particle creation in the TDHO}

{Let us now analyze the relation between the thermodynamical magnitudes associated to the TDHO}. The conventional definition of the quantum heat ($Q$) and work ($W$)  are \cite{Alicki1979, Alicki2004, Gemmer2009}
\begin{eqnarray}\label{Qdef01}
Q(t) &=&  \int_{t_0}^t {\rm Tr}\left( \frac{\partial \hat{\rho}(t')}{\partial t'} \hat{H}(t') \right) dt' ,   \\ \label{Wdef01}
W(t) &=&   \int_{t_0}^t {\rm Tr}\left( \hat{\rho}(t') \frac{\partial \hat{H}(t')}{\partial t'}  \right) dt'  .
\end{eqnarray}
Together with the definition of the total energy,
\be\label{energy01}
E(t) = \langle \hat H(t) \rangle = {\rm Tr}(\hat \rho(t) \hat H(t)) ,
\ee
they automatically satisfy the first principle of thermodynamics. Under unitary evolution,
\be
\frac{\partial \hat\rho_S(t)}{\partial t} =  -i [H_S(t) , \rho_S(t)]  ,
\ee
so the heat in equation (\ref{Qdef01}) is $ Q = 0$  (i.e. $\Delta E = W$). Specifically, regarding the expression (\ref{Qdef01}), unitary evolution implies adiabaticity. {However, for a quantum closed system heat can also be defined in terms of the change of energy due to a redistribution of the population of the energy levels \cite{Deffner2008, Polkovnikov2008, Polkovnikov2011, Dalessio2016} and, as we have seen in the preceding section,  there is in general a redistribution of these populations in the non quasi-static evolution of the TDHO. Then, we need to extend the definition of heat to account both for open and closed systems.}

Furthermore, the derivation of the definitions of heat and work (\ref{Qdef01}-\ref{Wdef01}) from that of the total energy \eqref{energy01} is not unique. For example, as it is pointed out in Ref. \cite{RP2011b}, one has
\be\label{energy01b}
E(t) = {\rm Tr}( \rho(t)  H(t)) = {\rm Tr}(\tilde \rho(t) \tilde H(t)) ,
\ee
provided that $\tilde \rho$ and $\tilde H$ are related to $\rho$ and $H$ by  a unitary transformation. Equation \eqref{energy01b} is satisfied, but it is not generally true that the associated definitions of heat and work are the same, i.e. $Q \neq \tilde Q$ and $W \neq \tilde W$. {As we have already pointed out, the issue of the basis becomes important \cite{Dalessio2016} (see, also, Refs. \cite{Calabrese2007, Foini2017}). In thermodynamical terms, it seems natural \cite{Polkovnikov2008, Rigol2008, Polkovnikov2011, Dalessio2016} to work in the instantaneous energy basis. Then, following the principles given in Ref. \cite{Polkovnikov2008}, we will exploit the invariance \eqref{energy01b} to look for a re-definition of the quantum heat that also accounts for the creation of particles in the non static regime of  the unitary evolution of the TDHO.}
Let us first note that for an initial number state $|N\rangle$, the time-dependent energy of the TDHO \eqref{energy01} can be written as
\be\label{energy02}
E(t) =  \omega(t) \left(  N_\omega(t) +\frac{1}{2} \right),
\ee
where
\be\label{MVNomega01}
N_\omega(t) \equiv \langle \hat N_\omega \rangle =   \left( |\alpha_\omega|^2 + |\beta_\omega|^2\right) N + |\beta_\omega|^2,
\ee
with {(see, Eqs. (\ref{alpha01}-\ref{beta01})}
\beq\label{modalphabeta01}
|\alpha_\omega|^2 + |\beta_\omega|^2 &=& \frac{\omega(t)}{2\omega_0} \left( \sigma^2  + \frac{\omega_0^2}{\sigma^2\omega^2(t)} + \frac{\dot\sigma^2}{\omega^2(t)} \right)  , \\ \label{modbeta01}
|\beta_\omega|^2 &=& \frac{\omega(t)}{4\omega_0} \left[ \left( \sigma - \frac{\omega_0}{\sigma \omega(t)} \right)^2 + \frac{\dot\sigma^2}{\omega^2(t)}  \right].
\eeq
The time variation of the energy \eqref{energy02} can then be written as
\be\label{Edotomega01_}
\dot E =  \dot\omega \left( N_\omega(t) + \frac{1}{2}\right) +  \omega \dot N_\omega(t) .
\ee 
In the quasi-static approximation,  $\frac{\dot\omega}{\omega} \ll 1$, and the change of the population of the energy levels is suppressed. In that approximation,
\be\label{quasistatic01}
\sigma^2 \approx \frac{\omega_0}{\omega(t)} \ , \ \frac{\dot\sigma}{\sigma} \propto \frac{\dot\omega}{\omega} \ll1 ,
\ee 
and then, 
\be
N_\omega(t)  \approx N + \frac{1}{2 \omega^2} \left( \frac{\dot\omega}{\omega} \right)^2 \left( N+\frac{1}{2}\right) ,
\ee
so up to first order in $\dot\omega/\omega$, $\dot N_\omega(t) \approx  0$, as it is assumed in some works  (see, for instance, Refs. \cite{Zhang2020, Picatoste2024, Oliveira2024b}). However, in the non quasi-static approximation $\dot N_\omega(t) \neq 0$ and both terms have to be considered in \eqref{Edotomega01_}. 

The first term in equation \eqref{Edotomega01_} accounts for the change in the total energy due to the explicit change of the frequency, {$\frac{\partial E}{\partial \omega} \dot\omega$}, and can be identified with the instantaneous work $\dot W_\omega$. The second term makes explicit the change in the energy caused by the redistribution of the population of the energy levels of the TDHO, {$\frac{\partial E}{\partial N_\omega} \dot N_\omega$}, and can be set with $\dot Q_\omega$. Using equations (\ref{MVNomega01}-\ref{modbeta01}), they read
\beq\label{newQ03}
\dot Q_\omega &=&   \frac{ \dot\omega}{2\omega_0} \left( \sigma^2 \omega - \frac{\omega_0^2}{\sigma^2 \omega} - \frac{\dot\sigma^2}{\omega}\right) \left( N +\frac{1}{2}\right) , \\ \label{newW03}
\dot W_\omega &=& \frac{ \dot\omega}{2\omega_0} \left( \sigma^2 \omega + \frac{\omega_0^2}{\sigma^2 \omega} + \frac{\dot\sigma^2}{\omega}\right) \left( N +\frac{1}{2}\right) ,
\eeq

\begin{figure}

\includegraphics[width=7cm]{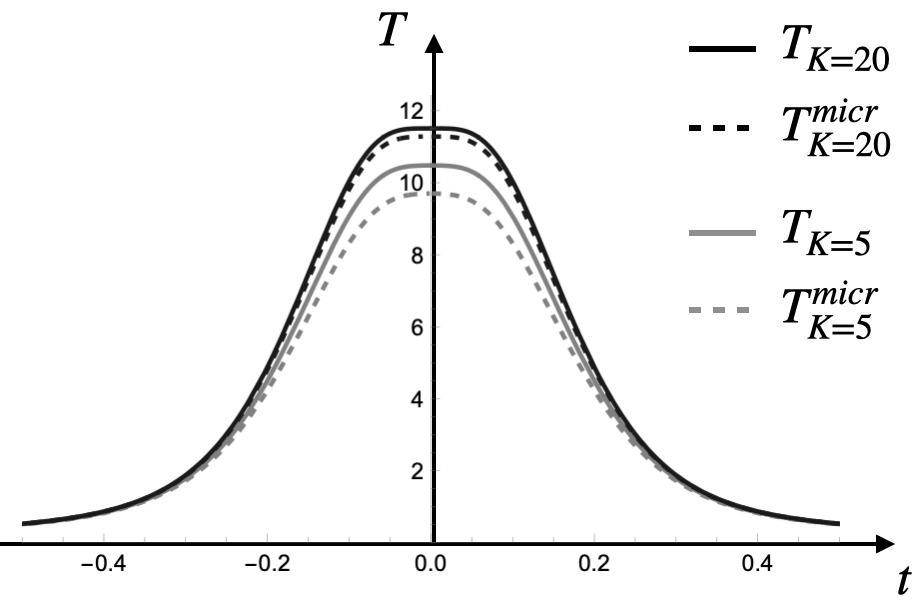}

\caption{Evolution of $T_{K}$ of the higher modes, given by \eqref{T_001}, versus $T_K^{micro}$, given by \eqref{Tmicro01}, which can be seen as a generalization of the former valid for all modes (not just the higher ones). For the higher modes, both approach the same value.}

\label{fig_Tmicr}

\end{figure}

\begin{figure}

\includegraphics[width=8.5cm]{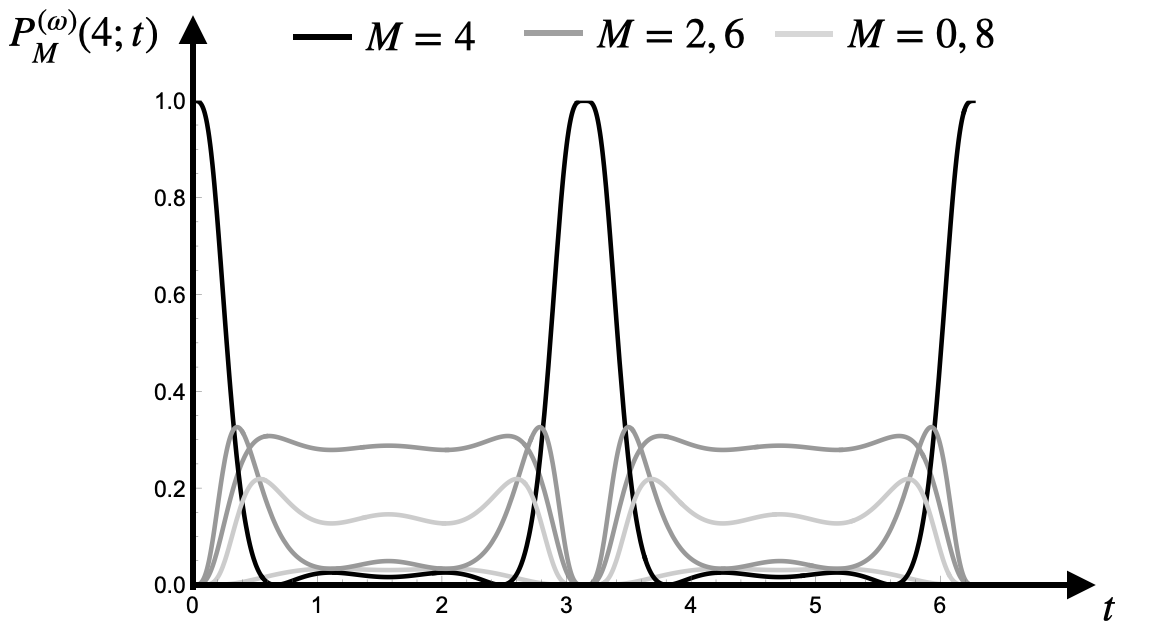}

\caption{Probability, $P_M^{(\omega)}(4;t)$, given by \eqref{ProbOme_01}, of finding the TDHO in the energy eigenstate $|M_\omega(t)\rangle$ when the initial state is the number state $|4\rangle$, for the Floquet frequency $\omega=\omega_3(t)$ of Fig.  \ref{fig_freq01}. The periods of energy state mixing and recovery to the original state follow one another periodically in a reversible way.}

\label{fig_PFloquet}

\end{figure}

\begin{figure}

\includegraphics[width=7.5cm]{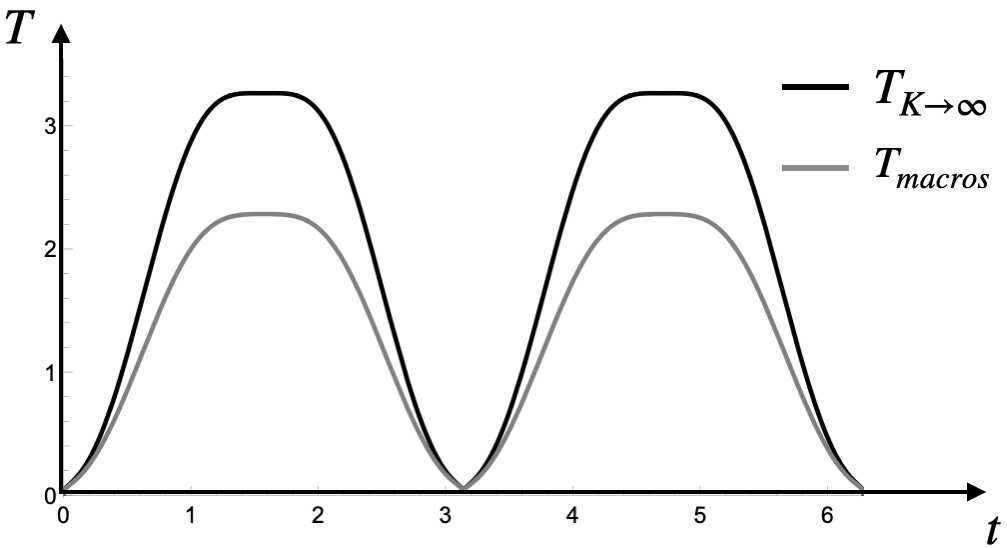}

\caption{Evolution of $T_{macr}$, given by \eqref{Tmacr01}, vs. $T_{K\rightarrow \infty}$ of the higher modes, given by \eqref{T_001}, for the Floquet frequency $\omega=\omega_3(t)$ of Fig.  \ref{fig_freq01}.  Temperatures periodically return to their initial value, $T=0$, over finite periods of time along the evolution of the TDHO.}

\label{fig_TFloquet}

\end{figure}

These values of the heat and work can also be obtained from the expressions \eqref{Qdef01} and \eqref{Wdef01}, but with a different representation of the density matrix and the Hamiltonian, given by 
\be\label{rhotilde01}
\tilde{\rho}_H(t) = \mathcal S_{0,\omega}  \rho_H \mathcal S_{0,\omega}^\dag ,
\ee
and
\be\label{Htilde01}
\tilde H_H =  \mathcal S_{0,\omega} H_H \mathcal S_{0,\omega}^\dag = \omega \left( \hat a^\dag_{0,S} \hat a_{0,S} + \frac{1}{2}\right)
\ee
where $\hat a_{0,S}^\dag$ and $\hat a_{0,S}$ are the constant creation and annihilation operators of the harmonic oscillator at time $t=0$, and
\be\label{S0o01}
\mathcal S_{0,\omega} = S_S\left(\log\sqrt{\frac{\omega_0}{\omega(t)}}\right) \, \mathcal U(t) ,
\ee
with the action of $S_S$ and $\mathcal U$ defined in App. \ref{app01}.  The time dependence of the Hamiltonian \eqref{Htilde01} is just contained in the frequency $\omega(t)$ because the operators $\hat a_{0,S}^\dag$ and $\hat a_{0,S}$ are constant operators, and $\tilde{\rho}_H(t)$ contains an explicit dependence on time (in the Heisenberg picture) because it represents the change in the populations of the energy levels along the evolution of the TDHO. By using Eq. \eqref{energy01b}, one can check that
\be\label{Etildedot01}
\dot E = {\rm Tr}( \tilde\rho_H \dot{\tilde{H}}_H ) + {\rm Tr} (\dot{\tilde{\rho}}_H \tilde H_H) ,
\ee
where the first term corresponds to $\dot W_\omega$ because
\beq\nn
{\rm Tr}( \tilde\rho_H \dot{\tilde{H}}_H )  &=& {\rm Tr}\left[ \tilde\rho_H  \dot\omega \left( \hat a^\dag_{0,S} \hat a_{0,S} + \frac{1}{2}\right) \right]  \\ \nn
&=&  \dot\omega  {\rm Tr}\left[  \rho_H \mathcal S_{0,\omega}^\dag \left( \hat a^\dag_{0,S} \hat a_{0,S} + \frac{1}{2}\right) \mathcal S_{0,\omega}  \right]  \\ 
&=&  \dot\omega  \left( N_\omega(t) + \frac{1}{2}\right)   = \dot W_\omega ,
\eeq
and the second term in Eq. \eqref{Etildedot01} can mainly be written as 
\beq\nn
 {\rm Tr} (\dot{\tilde{\rho}}_H \tilde H_H) &=& {\rm Tr} (\dot \rho_H H_H) + {\rm Tr} \left( \tilde{\rho}_H [\tilde{H}_H , \dot{\mathcal S}_{0,\omega} \mathcal S_{0,\omega}^\dag ] \right) \\  \label{Qgen03}
  &=& \dot Q_{nu} + \dot Q_\omega
\eeq
The first term in equation (\ref{Qgen03}) is zero if the evolution is unitary; it corresponds to the ordinary definition of heat and can be associated with the change in energy due to the change in the occupation number caused by dissipative effects. The second term can be seen as the quantum heat associated with the creation of particles along the unitary evolution in the non-quasi-static regime. It can be checked (see App. \ref{AppHeat01}) that it corresponds to \eqref{newQ03}. { Thus, the definition \eqref{Qgen03} extends the customary definition of heat to both closed and open systems.

\begin{figure}

\includegraphics[width=7cm]{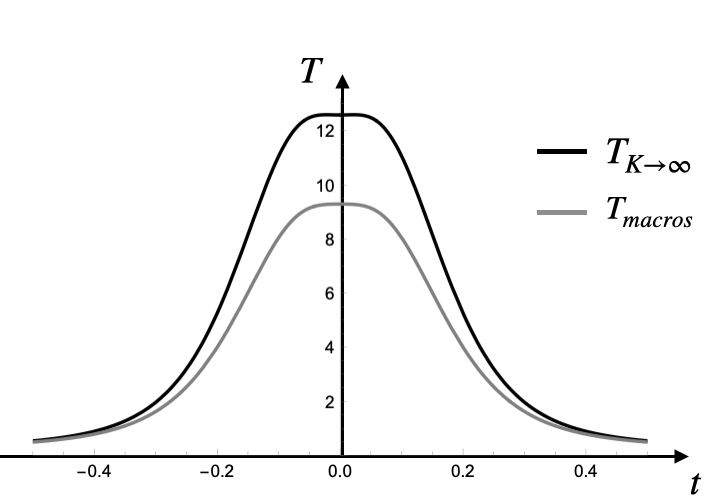}

\caption{Evolution of $T_{macro}$, given by \eqref{Tmacr01}, vs. $T_{K\rightarrow \infty}$ of the higher modes, given by \eqref{T_001}.  The former is smaller because it is the averaged over all modes (cold and hot modes) while the latter only refers to the largest (hottest) modes. In both cases, the effect is only relevant along the non quasi-static central region and the effect disappears after the central quench.}

\label{fig_Tmacr}

\end{figure}

On the other hand, heat is intimately connected to entropy and the second law of thermodynamics \cite{Polkovnikov2008, Polkovnikov2011, Dalessio2016}. However, the von Neumann entropy is zero for a closed system that undergoes unitary evolution. Given the close relationship between heat and entropy it  seems reasonable  to assume a different definition of the entropy too. Furthermore, von Neumann entropy is a good measure of the lack of knowledge of the state of the system but is not actually what is measured in thermodynamical experiments \cite{Deutsch2018, Oliveira2024b}, which is rather the probability $P_K$ of measuring the system on the $K$-energy level. These probabilities are given by the diagonal elements of the density matrix so instead of von Neumann entropy one can consider the \emph{diagonal entropy} \cite{Polkovnikov2008, Polkovnikov2011, Santos2011, Dalessio2016, Oliveira2024b}
\be\label{Dentropy01}
S_d = - \sum_K P_K \log P_K , 
\ee
where  in the case of a TDHO that evolves from the initial state $|N\rangle$, the probability $ P_K(N;t)$ is given by \eqref{ProbOme_01}. It can be shown \cite{Dalessio2016} that the diagonal entropy satisfies the fundamental thermodynamic relation 
\be
dE = T dS_d - F_\lambda d\lambda ,
\ee
where in the case of the TDHO, $\lambda = \omega$, and
\be
F_\omega = - \left. \frac{\partial E(\omega)}{\partial \omega}\right|_{S_d} .
\ee
For the unitary (reversible) evolution of the TDHO these definitions are consistent with \eqref{Etildedot01}, with $\dot Q_\omega$ and $\dot W_\omega$ given by \eqref{newQ03} and \eqref{newW03}, respectively, and 
\be\label{Tther01}
T \, \dot S_d = \dot Q_\omega .
\ee
From the microscopic point of view, $\dot Q_\omega$ represents the heating of the system caused by transitions between levels, \emph{independent of whether those transition are induced by contact with another system of by changing non-adiabatically some coupling $\lambda$, or both} (cfr. Ref. \cite{Dalessio2016}, p.63). From \eqref{Tther01}, we obtain then another  definition of an effective macroscopic temperature, given by
\be\label{Tmacr01}
T_{macro}(t) = \frac{\dot Q_\omega}{\dot S_d }  = \omega(t) \, \frac{\dot N_\omega}{\dot S_d } 
\ee
where $\dot N_\omega$ and $\dot S_d $ are obtained from the time derivatives of \eqref{MVNomega01} and \eqref{Dentropy01}, respectively. It is represented in Fig. \ref{fig_Tmacr}, where it can be verified that the temperature $T_{macro}(t)$ is smaller than $T_{K\gg1}(t)$, given in \eqref{T_001}, since the latter only accounts for the energy change due to the highest (hottest) modes while $T_{macro}(t)$ is an average of the energy change due to transitions to all energy levels, including the lowest (coldest) ones. This can also be seen by looking at the values of $\dot S_d$ and $\dot N_\omega$ in terms of the probabilities $P_K$. Taking into account the condition, $\sum_K \dot P_K = 0$, they can be written as
\be
\dot S_d =  -\sum_K \dot P_K \log P_K \equiv \sum_K \dot S_K ,
\ee
and
\be
\dot N_\omega = \sum_K K \, \dot P_K \equiv \sum_K \dot N_K .
\ee
In that case, one can define a temperature-like parameter for each mode (not only for the largest ones), given by
\be\label{Tmicro01}
T_K^{micro}(t) = \omega \frac{\dot N_K}{\dot S_K} = - \frac{\omega\, K}{\log P_K} .
\ee
It can be checked that for the largest modes $T_K^{micro}(t) $ approaches the temperature of the higher modes given by \eqref{T_001} (see, Fig. \ref{fig_Tmicr}), and that 
\be\label{Tmacromicro01}
T_{macro}= \frac{\sum_K T_K^{micro} \dot P_K  \log P_K  }{\sum_K  \dot P_K  \log P_K} .
\ee
Then, $T_{macr}$ is the average value of $T_K^{micr}$ over all modes. It is also worth noticing that under the strong ETH, where the probability of all modes follows the Gibbs distribution at temperature $T$,
\be
 T_K^{micro} = T,  \ \forall K \ \Rightarrow \ T_{macro} = T \ .
\ee

The qualitative behavior of $T_{macro}$ is similar to that of the temperature of the higher modes. The effect is only significant in the non quasi-static regime of the frequency and returns to its original value $T\rightarrow 0$ in the quasi-static one\footnote{The limit $T\rightarrow 0$ is also considered the adiabatic limit in Ref. \cite{Garriga2005}.}. This thermalization and \emph{de-thermalization} effect would be specially interesting to study in periodically driven systems \cite{Grifoni1998, Kinoshita2006, Langen2015, Lazarides2014a, Lazarides2014b}, which in the case of the TDHO can be represented by a Floquet frequency like the frequency $\omega_3(t)$ of \eqref{freq03} (see also Fig. \ref{fig_freq01}). In that case, the non-adiabatic effects are distributed periodically and the temperature returns to zero in a finite period of time over each period of the frequency (see Figs. \ref{fig_PFloquet}-\ref{fig_TFloquet}).

Let us finally make some comments on the experimental feasibility of the results obtained in this work. Clearly, monitoring a unitary evolution of an isolated quantum system is an experimental challenge. However, there have been significant experimental advances in recent years \cite{Greiner2002, Kinoshita2006, Bloch2008, Polkovnikov2011, Gring2012, Langen2015, Kaufman2016, Foini2017, Wang2022} and there are already experiments both to monitor the dynamic evolution of thermodynamic magnitudes associated with certain quantum systems \cite{Greiner2002, Kinoshita2006, Gangardt2008, Bloch2008} and to experimentally implement quantum systems that can be considered effectively isolated and evolving in a unitary way \cite{Gring2012, Langen2015, Kaufman2016}, at least over a significant timescale. The technique of \emph{time-of-flight} imaging \cite{Gangardt2008, Bloch2008, Polkovnikov2011} allows us to study the real-time evolution of the occupation of numbers states as well as their higher-order statistical moments \cite{Gangardt2008}. The results obtained in this work could presumably be accessible from similar experiments.

}

\section{Conclusions}

We have analyzed the evolution of the quantum state of a TDHO in the non-quasi-static regime. We have shown that there is, in general, a change in the population of the energy levels, and we have calculated the transition probabilities from an initial number state to the energy eigenstates of the harmonic oscillator along its entire evolution. It allows us to monitor the evolution of the energy levels in regions that are not fully considered in previous works, i.e. the transition and the non-quasi-static regions.

In the non-quasi-static regime of the frequency, the largest modes of the energy levels approach a thermal distribution with a temperature that does not depend on the specific conditions of the environment. In particular, it is not related to the temperature of any external thermal bath, but it spontaneously emerges from the evolution of the TDHO when the initial state is a pure number state. When the initial state is already in a thermal state at temperature $T_0$, the effect of the parametric amplification of the TDHO is to modify the value of the temperature along the evolution. We have calculated the first-order correction to that change in the non-quasi-static approximation.

The off-diagonal correlations between energy eigenstates turn out to be rapidly oscillating functions, so it is expected that they are easily destroyed in a real experiment by any dissipative of non unitary effect. However, if they can be preserved, or at least from a theoretical point of view, the spontaneous process of thermalization of the largest modes of the energy eigenstates is reversible under unitary evolution, and the TDHO recovers its original state when it reenters a quasi-static region after the non-quasi-static region. When the initial state is the vacuum, which can be seen as the limiting value of a thermal distribution with zero temperature, the population of the largest modes of the energy basis follow a thermal distribution with a temperature $T > 0$, after which, the distribution returns to its original vacuum state, at $T=0$. 
{We have extended the standard definitions of work and heat to account as well for change in the energy due to the redistribution of the populations of the energy levels along the unitary evolution of the TDHO. The definition is valid for both closed and open systems and it reduces to the customary one if the evolution of the TDHO is quasi-static. We have analyzed the relationship between the heat and the diagonal entropy and from the second law of thermodynamics we have derived an effective temperature for the whole ensemble. The qualitative behavior is similar to the microscopic temperature of the higher modes, it becomes significantly different from zero only in the non quasi-static region of the frequency of the TDHO and returns to its original value ($T\rightarrow 0$) when entering in a new quasi-static one. This effect is especially interesting in periodically driven systems where quasi-static and non-quasi-static regions succeed one another periodically. Thus, the effective temperatures of the TDHO return to the initial value (T=0) in a finite time periodically.}

We have revised the standard definitions of the quantum analogues of heat and work. Unlike the value of the total energy, which is invariant under a change of the representation of the density matrix and Hamiltonian operators, the values of heat and work depend on the representation used for these two operators. We have found a representation in terms of which the quantum heat preserves its interpretation as the change in the total energy caused by the redistribution of the populations of the energy levels, irrespective of whether this is accomplished by unitary (in the non-quasi-static regime) or non-unitary effects. In the unitary evolution of the TDHO, the changes produced in the heat and work are both eventually originated in the time dependence of the frequency. We have separated the effect that such dependence produces on the energy of the levels of the TDHO, and on the population of those levels. The former has been associated with the definition of work and the definition of heat. 

{The results presented can be used for new developments. This thermodynamic approach enables us to address responses to experimental challenges by estimating irreversibility. The essential role of this also appears in the QZE. Certainly, the thermodynamic considerations of quantum computing are interesting. In fact, it is assumed that quantum computing is, in principle, reversible. As Lucia has pointed out \cite{Lucia2023a}, new challenges of thermodynamics could be a topic for research for the optimization of quantum circuits. Undoubtedly, the QZE has drawn interest in the thermodynamic study of quantum information}. 


\section*{Acknowledgments}

The work of SJRP was supported by the Grant PID2021-123226NB-I00 (funded by MCIN/AEI/10.13039/501100011033 and by “ERDF A way of making Europe”). Funding for APC: Universidad Carlos III de Madrid  (Agreement CRUE-Madroño 2026)

\appendix

\section{Solution of the Schrödinger equation of the TDHO and evolution operator}\label{app01}

Let us consider the following operators
\be\label{AppSD01}
S(\beta) = e^{\frac{i}{2} \beta (\hat x \hat p_x + \hat p_x \hat x)}  \ , \ D(\alpha) = e^{-\frac{i}{2} \alpha \hat x^2} ,
\ee
and with them construct the unitary transformation $ U_0$ defined as
\be\label{frequency1operator01}
 U_0 =  S(-\log\sigma) D(-\sigma\dot\sigma) ,
\ee
where $\sigma(t)$ is the solution of equation \eqref{sigma01}. Then, $U_0$ relates the Hamiltonian of the TDHO \eqref{ham01a} with the Hamiltonian of the harmonic oscillator with constant frequency $\omega_0$, i.e.
\be
\hat H_{x,S} = \frac{1}{\sigma^2}  U_0 \hat H_{x,S}^{(0)}   U_0^\dag + i  \, \dot{ U}_0  U_0^\dag 
\ee
where
\be
\hat H_{x,S}^{(0)}  =  \frac{1}{2}  \left(\hat p_{x, S}^2 +  \omega_0^2 \, \hat x_S^2 \right)
\ee
Then, if $\psi(x,t)$ is the solution of the Schrödinger equation \eqref{sch01},
\be
\psi(x,t) =  U_0 \psi^{(0)}(x,t)
\ee
where $\psi^{(0)}(x,t)$ is the solution of the Schrödinger equation
\be
i \frac{\partial}{\partial \tau} \psi^{(0)}(x,\tau ) = \hat H_{x,S}^{(0)}   \psi^{(0)}(x,\tau) ,
\ee
with, $\dot \tau = \sigma^{-2}$, and the set of solutions
\be\label{U0t01}
\psi^{(0)}_N(x,\tau) = e^{-i \tau \hat H_{x,S}^{(0)}} \psi_N^{(0)}(x)  = e^{-i \left( N +\frac{1}{2}\right) \omega_0\tau(t)}  \psi_N^{(0)}(x)  ,
\ee
and,
\be\label{WFInv00b}
\psi_N^{(0)}(x) =  \frac{1}{\sqrt{2^N N! }} \left( \frac{\omega_0}{\pi}\right)^\frac{1}{4}  e^{-\frac{ \omega_0 x^2}{2}} \text{H}_N(\sqrt{\omega_0}x)
\ee
with $\text{H}_N$  the Hermite polynomial of order $N$. Now, using equations (\ref{U0t01}-\ref{WFInv00b}) and the action of $S$ and $D$ on a wave function $\psi(x)$ (see \cite{Seleznyova1995})
\beq
D(\alpha)  \psi(x) &=&  e^{-\frac{i}{2} \alpha  x^2}  \psi(x)   \\
S(\beta)  \psi(x) &=& e^{\frac{\beta}{2}} \psi(e^{\beta} \, x)
\eeq
one can check that the action of $\mathcal U$, 
\be\label{Ut01}
\mathcal U(t) =  e^{-\frac{i}{2} \log \sigma (\hat x \hat p_x + \hat p_x \hat x)} e^{\frac{i}{2} \sigma\dot\sigma\hat x^2} e^{-\frac{i}{2} \tau (\hat p_x^2+\omega_0^2 \hat x^2)} ,
\ee
on the wave function $\psi_N^{(0)}(x)$ gives  the wave function \eqref{WFInv01}.

\section{Calculation of the matrix elements $\langle M(t) | N_\omega\rangle$}\label{MElements01}

Under the customary scalar product of quantum mechanics,
\be
\langle \psi(x) | \phi(x) \rangle = \int_{-\infty}^\infty dx \, \bar\psi(x) \phi(x) 
\ee
the matrix element $\langle M(t) | N_\omega\rangle$ can be written as
\beq\nn
\langle M(t) | N_\omega \rangle &=& \int_{-\infty}^\infty dx \, \bar\psi_M \psi_N^{(\omega)} \\ \label{Melement01}
&=& \frac{e^{i(M+\frac{1}{2}) \omega_0 \tau(t)}}{\sqrt{2^{M+N} N! M!  \pi}} \left(\frac{\omega\omega_0}{\sigma^2}\right)^\frac{1}{4} \, I_{MN}(t)
\eeq
where, $\omega=\omega(t)$, and $\psi_M$ and $\psi_N^{(\omega)}$ are given by equations \eqref{WFInv01} and \eqref{WFOme01}, respectively, and $I_{MN}(t)$ is the integral of Hermite polynomials
\be\label{Int01}
I_{MN}(t) = \int_{-\infty}^\infty dx \, e^{- \phi(t) x^2} H_M\left( \frac{\sqrt{\omega_0} x}{\sigma}\right) H_N\left( \sqrt{\omega}\, x\right)
\ee
with
\be\label{phi01}
\phi(t) = \frac{\omega}{2\sigma} \left( \sigma + \frac{\omega_0}{\sigma \omega} + \frac{i\dot\sigma}{\omega} \right)  = \frac{\sqrt{\omega_0 \, \omega}}{\sigma} \alpha_\omega e^{+i\omega_0\tau}
\ee
and $\alpha_\omega$ given by Eq.\eqref{alpha01}. We can make the change $u = \sqrt{\phi} x$, and write the integral \eqref{Int01} as
\be\label{Int02}
I_{MN}(t) = \frac{1}{\sqrt{\phi}} \int_{-\infty}^\infty du \, e^{- u^2} H_M(a u) H_N(bu) = \frac{1}{\sqrt{\phi}} I_{MN}^{ab}
\ee
with, $a = \frac{\sqrt{\omega_0}}{\sigma\sqrt{\phi}}$ and $b = \sqrt{\frac{\omega}{\phi}}$, which satisfy
\beq \label{cond05}
\frac{1}{a^2} + \frac{1}{b^2} -\frac{1}{a^2 b^2} &=& |\alpha_\omega|^2 ,
 \eeq
with $\alpha_\omega$ given by equation \eqref{alpha01}. If $M+N$ is  even and $M\geq N$, the integral $I_{MN}^{ab}$ in Eq. \eqref{Int02} can be written as (see Eq. (2.2) of Ref. \cite{Bailey1948})
\beq\nn
I_{MN}^{ab}  = \frac{M! (2ab)^N (a^2-1)^\frac{M-N}{2} \pi^\frac{1}{2}}{\left(\frac{M-N}{2}\right)!} \\ \label{IMN02}F\left[ -\frac{N}{2} , \frac{1-N}{2} ; 1+\frac{M-N}{2}; - |\beta_\omega|^2  \right]
\eeq
where $|\beta_\omega|^2 = |\alpha_\omega|^2 - 1$. For $M < N$, the same procedure applies by changing $N \leftrightarrow M$ and $a \leftrightarrow b$, so for $N>M$
\beq\nn
I_{MN}^{ab}  = \frac{N! (2ab)^M (b^2-1)^\frac{N-M}{2} \pi^\frac{1}{2}}{\left(\frac{N-M}{2}\right)!} \\ \label{IMN03}F\left[ -\frac{M}{2} , \frac{1-M}{2} ; 1+\frac{N-M}{2}; - |\beta_\omega|^2  \right]
\eeq

Using the expansions of the associated Legendre function in terms of hypergeometric functions given in Ref. \cite{Bateman1953}, in particular, equations (9) and (24) of Ref. \cite{Bateman1953},
\beq\nn
P^\frac{N-M}{2}_\frac{N+M}{2}(z) = \frac{2^\frac{N-M}{2} (z^2-1)^\frac{M-N}{4} z^N}{\Gamma\left(1+\frac{M-N}{2} \right)} \\ F\left(-\frac{N}{2}, \frac{1-N}{2}; 1+\frac{M-N}{2}; 1-\frac{1}{z^2}\right)
\eeq
one can write,
\beq\nn
F\left(-\frac{N}{2} , \frac{1-N}{2} ; 1 + \frac{M-N}{2}; \right.  &  -|\beta_\omega|^2 & \left.  \right) = \Gamma\left(1+\frac{M-N}{2} \right) \\ \label{eq01}  2^\frac{M-N}{2}  \left( -\frac{|\alpha_\omega|}{|\beta_\omega|}\right)^\frac{M-N}{2} & |\alpha_\omega|^N & P^\frac{N-M}{2}_\frac{N+M}{2}\left(\frac{1}{|\alpha_\omega|}\right)
\eeq
and obtain for $M \geq N$
\be\label{Int03}
I_{MN}^{ab}  = \frac{M! \pi^\frac{1}{2}}{2^{-\frac{M+N}{2}}} e^{-i(\omega_0\tau M + \varphi_0(t)) } P^\frac{N-M}{2}_\frac{N+M}{2}\left(\frac{1}{|\alpha_\omega|}\right)
\ee
where,
\be
\varphi_0(t) = \frac{\theta_\alpha}{2}(N+M)-\frac{\theta_\beta}{2}(M-N)
\ee
with $\theta_\alpha$ and $\theta_\beta$  the phases of $\alpha_\omega$ and $\beta_\omega$, respectively,  $\alpha_\omega =|\alpha_\omega| e^{i\theta_\alpha}$ and $\beta_\omega =|\beta_\omega| e^{i\theta_\beta}$. Combining equations \eqref{Melement01} with (\ref{phi01}-\ref{Int02}) and  \eqref{Int03}, one gets
\beq\nn
\langle M(t) | N_\omega \rangle =  \sqrt{\frac{M!}{N!}} e^{-\frac{i\theta_\alpha}{2}(N+M+ 1)+\frac{i\theta_\beta}{2}(M-N) } \\ \label{Melement02}  \frac{1}{\sqrt{|\alpha_\omega|}}P^\frac{N-M}{2}_\frac{N+M}{2}\left(\frac{1}{|\alpha_\omega|}\right)
\eeq
which essentially coincides\footnote{The phases are slightly different because Brown does not use the proper diagonal representation but some approximation to the the quasi-static solution that is valid only in the two asymptotic regions that they consider in the remote past and future. Our solution is more general because it can be applied to any value of the frequency (quasi-static or not) and at any moment of time (not only in the remote past and future).} with the result obtained by Ref. \cite{Brown1979}, Eq. (5.17).

Following the same steps, it is obtained for $N > M$ (be aware that now it is not exactly changing $N \leftrightarrow M$, the phase $e^{-i\tau M}$ remains being the same so it doesn't change to $e^{-i\tau N}$, and there is also a alternating negative sign)
\be\label{Int04}
I_{MN}^{ab}  = \frac{N! \pi^\frac{1}{2} (-1)^{\frac{N-M}{2}}}{2^{-\frac{M+N}{2}}} e^{-i(\omega_0\tau M + \varphi_0(t)) } P^\frac{M-N}{2}_\frac{M+N}{2}\left(\frac{1}{|\alpha_0|}\right)
\ee
and thus  (for $N > M$)
\beq\nn
\langle M(t) | N_\omega \rangle =  \sqrt{\frac{N!}{M!}} e^{-\frac{i\theta_\alpha}{2}(N+M+ 1)-\frac{i\theta_\beta}{2}(N-M) } \\ \label{Melement03}  \frac{(-1)^\frac{N-M}{2} }{\sqrt{|\alpha_\omega|}}P^\frac{M-N}{2}_\frac{M+N}{2}\left(\frac{1}{|\alpha_\omega|}\right) .
\eeq
However, knowing that $\frac{N\pm M}{2}$ is an integer and that the argument of the associated Legendre function is a real number, then, we can use Eq. (8.752.2) of Ref. \cite{Gradshteyn2007}, to transform equation \eqref{Melement02} into \eqref{Melement03}. In that case, it turns out that both equations \eqref{Melement02} and \eqref{Melement03} are valid in the two cases, $M <N$ and $M \geq N$.

\section{Calculation of the quantum heat  $\dot Q_\omega(t)$}\label{AppHeat01}

From equation \eqref{S0o01}, one has
\beq\nn
\dot S_{0,\omega} \, S_{0,\omega}^\dag  &=&  \dot S_S \, S_S^\dag   -i \,  S_S \, H_S \, S_S^\dag , \\ 
&=& -\frac{i}{4} \frac{\dot\omega}{\omega} (\hat x_S \hat p_{x,S} + \hat p_{x,S} \hat x_S) - i   \tilde H_H
\eeq
where the operator $S_S \equiv S_S(\frac{1}{2}\log\frac{\omega_0}{\omega})$ is defined in appendix \eqref{AppSD01}, we have used $\dot{\mathcal U}\mathcal U^\dag = -i H_S$, and $\tilde H_H$ is given by Eq. \eqref{Htilde01}. Now,
\beq\nn
[\tilde H_H, \dot S_{0,\omega} \, S_{0,\omega}^\dag]  &=&  -\frac{i}{4} \frac{\dot\omega}{\omega}  [ \tilde H_H,\hat x_S \hat p_{x,S} + \hat p_{x,S} \hat x_S] \\
&=& -\frac{\dot\omega}{2\omega_0} \left( \hat p_{x,S}^2 - \omega_0 \hat x_S^2\right) \\
&=& \frac{\dot \omega}{2} \left( (\hat a_{0,S}^\dag)^2 + \hat a_{0,S}^2 \right) ,
\eeq
where  $\hat a_{0,S}^\dag$ and $\hat a_{0,S}$ are the constant creation and annihilation operators of the harmonic oscillator at time $t=0$. Then,
\beq\nn
{\rm Tr} \left( \tilde\rho_H [\tilde H_H, \dot S_{0,\omega} \, S_{0,\omega}^\dag] \right) &=& \frac{\dot\omega}{2} {\rm Tr} \left[\tilde\rho_H \left( (\hat a_{0,S}^\dag)^2 + \hat a_{0,S}^2 \right) \right] \\ \nn
&=& \frac{\dot\omega}{2} {\rm Tr} \left[\rho_H \left( (\hat a_{\omega,H}^\dag)^2 + \hat a_{\omega,H}^2 \right) \right] ,
\eeq
with
\beq\label{ao05}
\hat a_{\omega,H}(t) &=&  \alpha_\omega(t) \,  \hat a_{0,S} + \beta_{\omega}(t) \, \hat a_{0,S}^\dag  , \\ \label{ao06}
\hat a_{\omega,H}^\dag(t) &=&  \alpha_\omega^*(t) \,  \hat a_{0,S}^\dag + \beta_{\omega}^*(t) \, \hat a_{0,S}  .
\eeq
By using $\rho_H = |N\rangle\langle N|$, it is finally obtained
\beq\nn
{\rm Tr} \left( \tilde\rho_H [\tilde H_H, \dot S_{0,\omega} \, S_{0,\omega}^\dag] \right) &=& \frac{\dot\omega}{2} \langle N|  (\hat a_{\omega,H}^\dag)^2 + \hat a_{\omega,H}^2 |N\rangle  \\ \nn
&=& 2 \dot\omega {\rm Re} \left( \alpha_\omega \beta_\omega \right) \left( N+\frac{1}{2}\right) \\ \nn
&=& \frac{\dot\omega\omega}{2\omega_0} \left( \sigma^2 - \frac{\omega_0^2}{\sigma^2\omega^2} - \frac{\dot\sigma^2}{\omega^2}\right) \left( N+\frac{1}{2}\right) \\
&=&  \dot Q_\omega(t) .
\eeq


%


\end{document}